\documentclass[preprintnumbers,amsmath,nofootinbib,secnumarabic]{revtex4}
\usepackage{graphicx}
\usepackage{mathrsfs}
\usepackage{dsfont}
\usepackage{dcolumn,comment}
\usepackage{color}
\usepackage{amssymb,amsmath,amsbsy,bbold}
\usepackage{latexsym}
\numberwithin{equation}{section}

\def\beq{\begin{equation}}
\def\eeq{\end{equation}}
\newenvironment{Eqnarray}%
     {\arraycolsep 0.14em\begin{eqnarray}}{\end{eqnarray}}
\def\beqa{\begin{Eqnarray}}
\def\eeqa{\end{Eqnarray}}
\def\bea{\begin{Eqnarray*}}
\def\eea{\end{Eqnarray*}}
\renewcommand{\Re}{{\rm Re}}
\renewcommand{\Im}{{\rm Im}}
\def\lsim{\mathrel{\raise.3ex\hbox{$<$\kern-.75em\lower1ex\hbox{$\sim$}}}}
\def\gsim{\mathrel{\raise.3ex\hbox{$>$\kern-.75em\lower1ex\hbox{$\sim$}}}}

\def\abar{{\bar a}}
\def\bbar{{\bar b}}
\def\cbar{{\bar c}}
\def\dbar{{\bar d}}

\def\hp{{H^+}}
\def\hm{{H^-}}

\def\mc{m_{H^\pm}}

\def\phm{\phantom{-}}

\def\Ref#1{ref.~\cite{#1}}
\def\refs#1#2{refs.~\cite{#1} and \cite{#2}}

\def\eq#1{eq.~(\ref{#1})}
\def\eqs#1#2{eqs.~(\ref{#1}) and (\ref{#2})}

\def\ifmath#1{\relax\ifmmode #1\else $#1$\fi}
\def\ls#1{\ifmath{_{\lower1.5pt\hbox{$\scriptstyle #1$}}}}
\def\lss#1{\ifmath{^{\,\lower2.5pt\hbox{$\scriptstyle #1$}}}}
\def\lsup#1{^{\lower 6pt\hbox{$\scriptstyle#1$}}}
\def\llsup#1{^{\lower 3pt\hbox{$\scriptstyle#1$}}}
\def\lasup#1{^{\lower 2pt\hbox{$\scriptstyle#1$}}}

\def\half{\tfrac{1}{2}}

\def\quarter{\tfrac{1}{4}}

\def\mw{{m_W^2}}
\def\mz{{m_Z^2}}

\def\zth{Z_3}
\def\zthf{Z_3+Z_4}
\def\zone{Z_1}
\def\zf{Z_4}

\def\zfii{\Im(Z_5 \,e^{-2i\theta_{23}})}

\def\lsup#1{^{\lower 6pt\hbox{$\scriptstyle#1$}}}
\def\phaa{\phantom{AA}}
\def\zsixr{\Re(Z_6 e^{-i \theta_{23}})}
\def\zsixi{\Im(Z_6 e^{-i \theta_{23}})}
\def\zfiver{\Re(Z_5 \,e^{-2i\theta_{23}})}
\def\ddel{\!\!\mathrel{\raise1.5ex\hbox{$\leftrightarrow$\kern-.85em
\lower1.7ex\hbox{$\partial$}}}}

\def\thet{e^{i \theta_{23}}}

\def\ctw{c_{2W}}

\def\mcs{m^2_{H^\pm}}

\def\PMzz{\delta \Pi_{ZZ}(\mz)}
\def\Pozz{\delta \Pi _{ZZ}(0)}
\def\PMpzz{\delta \Pi'_{ZZ}(\mz)}
\def\Popzz{\delta \Pi'_{ZZ}(0)}
\def\Popww{\delta \Pi'_{WW}(0)}
\def\Poaa{\delta \Pi_{\gamma\gamma}(0)}
\def\Poza{\delta \Pi_{Z\gamma}(0)}

\def\Poww{\delta \Pi _{WW}(0)}
\def\Pww{\delta \Pi _{WW}(\mw)}
\def\Ppww{\delta \Pi' _{WW}(\mw)}
\def\Paa{\delta \Pi_{\gamma\gamma}(\mz)}
\def\Pza{\delta \Pi_{Z\gamma}(\mz)}
\def\Popaa{\delta \Pi'_{\gamma\gamma}(0)}
\def\Popza{\delta \Pi'_{Z\gamma}(0)}

\def\QMzz{\delta \Pi_{ZZ}(q^2)}
\def\QMpzz{\delta \Pi'_{ZZ}(q^2)}
\def\Qpaa{\delta \Pi'_{\gamma\gamma}(q^2)}
\def\Qww{\delta \Pi _{WW}(q^2)}
\def\Qpww{\delta \Pi' _{WW}(q^2)}
\def\Qaa{\delta \Pi_{\gamma\gamma}(q^2)}
\def\Qza{\delta \Pi_{Z\gamma}(q^2)}
\begin{document}
\title{What the Oblique Parameters S, T, and U and Their Extensions Reveal About the 2HDM: A Numerical Analysis}

\author{Gerhardt Funk\footnote{gfunk@my.centenary.edu}}\affiliation{Department of Physics, Centenary College}
\author{Deva O'Neil\footnote{doneil@bridgewater.edu}}\affiliation{Department of Physics, Bridgewater College}
\author{R. Michael Winters\footnote{Raymond.Winters@mail.mcgill.ca}}\affiliation{Center for Interdisciplinary Research in Music Media and Technology, McGill University}

\begin{abstract}
The oblique parameters S, T, and U and their higher-order extensions (V, W, and X) are observables that combine electroweak precision data to quantify deviation from the Standard Model.  These parameters were calculated at one loop in the basis-independent CP-violating Two-Higgs Doublet Model (2HDM).  The scalar parameter space of the 2HDM was randomly sampled within limits imposed by unitarity and found to produce values of the oblique parameters within experimental bounds, with the exception of T.   The experimental limits on T were used to predict information about the mass of the charged Higgs boson and the difference in mass between the charged Higgs boson and the heaviest neutral Higgs boson.  In particular, it was found that the 2HDM predicts -600  GeV $ < \mc - m_3 < $100 GeV, with values of $\mc > 250$ GeV being preferred.  The mass scale of the new physics ($M_{NP}$) produced by random sampling was consistently fairly high, with the average of the scalar masses falling between 400 and 800 GeV for $Y_2 = m_W^2$, although the model can be tuned to produce a light neutral Higgs mass ($\sim 120$ GeV).  Hence, the values produced for V, W, and X fell well within .01 of zero, confirming the robustness of the linear expansion approximation. Taking the CP-conserving limit of the model was found to not significantly affect the values generated for the oblique parameters.  
\end{abstract}

\maketitle

\section{Introduction}
In supersymmetric models, the scalar sector of the theory features multiple Higgs bosons. In the minimal SUSY scenario, which requires two Higgs doublets, the phenomenology features three neutral Higgs bosons and one charged Higgs boson pair.  However, if multiple Higgs particles are discovered in the absence of evidence of squarks or other SUSY partners, it will not be immediately apparent what symmetries govern these phenomena.  Because a generic Two-Higgs-Doublet-Model (2HDM) produces unrealistically large flavor-changing neutral currents, naturalness arguments would lead one to expect that some of symmetry is present in the Higgs sector, but the symmetry need not be associated with SUSY.  In this paper we examine a generic 2HDM in a formalism that does not impose symmetries on the scalar sector, as in \refs{davidson}{haberoneil}.  Absent such symmetries, all physical observables must be basis-independent (invariant with respect to transformation in Higgs-flavor space).  

In the case that the scale of the new physics is not too much larger than the Z boson mass, the 2HDM produces non-zero shifts of the oblique parameters S, T, and U as well as the higher-order parameters V, W, and X.  In this paper we calculate the six parameters in the 2HDM and randomly sample the 2HDM parameter space to compare the predictions of the model to the experimental results. The goal of this work is to present basis-independent formulae for the extended parameters (V, W, and X) and to answer the following questions: How do the values generated in the 2HDM for the oblique parameters compare with the  experimental values?  Does the 2HDM predict significant shifts in the higher-order parameters? (ie, is any additional information provided by going beyond S, T, and U?)  What do these results for the oblique parameters reveal about the phenomenology (neutral and charged scalar masses) of the 2HDM?  

We also examine factors that may affect the validity of our results, such as the differences between the various definitions of the oblique parameters appearing in the literature, the effect of our choice of $Y_2$ (an unconstrained parameter in the scalar potential) on the size of the scalar masses, the errors inherent in using experimental limits from the PDG that use a fixed reference mass, and the extent to which taking the CP-conserving limit changes the numerical results.

\subsection{The Basis-Independent Two-Higgs Doublet Model}
The theory of a basis-independent CP-violating 2HDM has been developed in \refs{haberoneil}{haberoneil2}.  What follows is a brief summary of the elements of the theory that will be relevant for the calculations in this paper. 

The most generic scalar potential for a model of two Higgs doublets is
 \beq \label{genericpot}
\mathcal{V}=Y_{a\bbar}\Phi_\abar^\dagger\Phi_b
+\half Z_{a\bbar c\dbar}(\Phi_\abar^\dagger\Phi_b)
(\Phi_\cbar^\dagger\Phi_d)\,,
\eeq
where the Higgs doublet fields are written as $\Phi_a(x)\equiv (\Phi^+_a(x)\,,\,\Phi^0_a(x))$, with  $a=1,2$; complex conjugation converts a barred index to unbarred and vice versa.   The fields $\Phi_a$ can be redefined by an arbitary U(2) transformation which rotates them to a different basis in Higgs flavor space. The coefficients $Y_{a\bbar}$ and $Z_{a\bbar c\dbar}$ in \eq{genericpot} are likewise basis-dependent.  Any measurable parameters of the 2HDM must be invariant with respect to such a transformation.  Invariants can be constructed by summing over all Higgs flavor indices, pairing unbarred and barred indices, as is done in the covariant equation \ref{genericpot}. 

The vacuum expectation values of the Higgs doublets are
\beq \label{emvev}
\langle \Phi_a
\rangle={\frac{v}{\sqrt{2}}} \left(
\begin{array}{c} 0\\ \widehat v_a \end{array}\right)\,,
\eeq
where $v=246$~GeV and $\widehat v_a$ is a vector of unit norm. One can define new doublet fields as follows:
\beq \label{hbasisdef}
H_1=(H_1^+\,,\,H_1^0)\equiv \widehat v_{\abar}^*\Phi_a\,,\qquad\qquad
H_2=(H_2^+\,,\,H_2^0)\equiv  \epsilon_{\bbar\abar}
\widehat v_b\Phi_a \,.
\eeq
This basis choice is known as the Higgs basis.  In this basis, the 2HDM scalar potential becomes
\beqa \label{hbasispot}
\mathcal{V}&=& Y_1 H_1^\dagger H_1+ Y_2 H_2^\dagger H_2
+[Y_3 H_1^\dagger H_2+{\rm h.c.}]\nonumber \\[5pt]
&&\quad
+\half Z_1(H_1^\dagger H_1)^2 +\half Z_2(H_2^\dagger H_2)^2
+Z_3(H_1^\dagger H_1)(H_2^\dagger H_2)
+Z_4( H_1^\dagger H_2)(H_2^\dagger H_1) \nonumber \\[5pt]
&&\quad +\left\{\half Z_5 (H_1^\dagger H_2)^2
+\big[Z_6 (H_1^\dagger H_1)
+Z_7 (H_2^\dagger H_2)\big]
H_1^\dagger H_2+{\rm h.c.}\right\}\,.
\eeqa
 From the scalar potential in \eq{hbasispot} one can extract the squared-mass matrix for the three neutral states,
\beq \label{matrix33}
\mathcal{M}=v^2\left( \begin{array}{ccc}
\hspace{-.1 in}Z_1&\,\, \Re(Z_6) &\,\, -\Im(Z_6)\\
\hspace{-.1 in}\Re(Z_6)  &\,\, \half\left[Z_3+Z_4+\Re(Z_5)\right]+Y_2/v^2 & \,\,
- \half \Im(Z_5)\\\hspace{-.1 in} -\Im(Z_6) &\,\, - \half \Im(Z_5) &\,\,
 \half\left[Z_3+Z_4-\Re(Z_5)\right]+Y_2/v^2\end{array}\right).
\eeq
The field $H_2$ and the coefficients $Y_3$ and $Z_{5,6,7}$ depend on the choice of basis, and thus are not physically meaningful by themselves unless a symmetry (such as SUSY) is selecting a preferred basis.  Thus, the mass matrix in \eq{matrix33} is not invariant. This matrix is diagonalized with the transformation
\beq \label{rmrt}
R\mathcal{M} R^T=\mathcal{M}_D\equiv {\rm diag}~(m_1^2\,,\,m_2^2\,,\,m_3^2)\,,
\eeq
with $R$ defined as follows:\footnote{In the CP-conserving 2HDM, only a single mixing angle is required.   In this paper, the formalism of the CP-violating theory will be used, but as we will show, the numerical results for the oblique parameters do not change significantly in the CP-conserving limit.} 
\beqa \label{rmatrix}
R=R_{12}R_{13}R_{23} &=&\left( \begin{array}{ccc}
c_{12}\,\, &-s_{12}\quad &0\\
s_{12}\,\, &\phm c_{12}\quad &0\\
0\,\, &\phm 0\quad &1\end{array}\right)\left( \begin{array}{ccc}
c_{13}\quad &0\,\, &-s_{13}\\
0\quad & 1\,\,&\phm 0\\
s_{13}\quad &0\,\, &\phm c_{13}\end{array}\right) \left( \begin{array}{ccc}
1\quad &0\,\, &\phm 0\\
0\quad &c_{23}\,\, &-s_{23}\\
0\quad &s_{23}\,\, &\phm c_{23}\end{array}\right) \nonumber \\[10pt]
&=&
\left( \begin{array}{ccc}
c_{13}c_{12}\quad &-c_{23}s_{12}-c_{12}s_{13}s_{23}\quad &-c_{12}c_{23}s_{13}
+s_{12}s_{23}\\[6pt]
c_{13}s_{12}\quad &c_{12}c_{23}-s_{12}s_{13}s_{23}\quad
& -c_{23}s_{12}s_{13}-c_{12}s_{23}\\
s_{13}\quad &c_{13}s_{23}\quad &c_{13}c_{23}\end{array}\right)\,,
\eeqa
where $c_{ij}\equiv \cos\theta_{ij}$ and $s_{ij}\equiv\sin\theta_{ij}$.
The angles $\theta_{12}$ and $\theta_{13}$ are invariant; the third angle $\theta_{23}$ depends on the basis choice.  However, the product $\thet H_2$ is invariant.  Thus, one can express the squared-mass matrix entirely in terms of invariants: 
\beq \label{mtilmatrix}
\widetilde{\mathcal{M}}\equiv R_{23}\mathcal{M}R_{23}^T=\!
v^2\!\left( \begin{array}{ccc}
\hspace{-.12 in}Z_1&\,\, \Re(Z_6 \, e^{-i\theta_{23}}) &\,\, -\Im(Z_6 \, e^{-i\theta_{23}})\\
\hspace{-.12 in}\Re(Z_6 e^{-i\theta_{23}}) &\,\,\Re(Z_5 \,e^{-2i\theta_{23}})+ A^2/v^2 & \,\,
\hspace{-.12 in}- \half \Im(Z_5 \,e^{-2i\theta_{23}})\\ -\Im(Z_6 \,e^{-i\theta_{23}})
 &\,\, - \half \Im(Z_5\, e^{-2i\theta_{23}}) &\,\, A^2/v^2\end{array}\right)\!,
\eeq
where 
${A}^2$ is defined by
\beq \label{madef}
{A}^2\equiv Y_2+\half[Z_3+Z_4-\Re(Z_5 e^{-2i\theta_{23}})]v^2\,.
\eeq
In this paper we use a mass ordering of the neutral fields such that $m_1\leq m_2\leq m_3$. The remaining Higgs particle has the mass
\beq \mc = \sqrt{Y_2 + \half \zth v^2} \label{mch}.\eeq

In calculating the oblique parameters, certain invariant combinations of the mixing angles appear.  These combinations are presented in Table \ref{tab1}. 
\begin{table}[h!]
\centering
\caption{The U(2)-invariant quantities $q_{k\ell}$ are functions of the
the neutral Higgs mixing angles $\theta_{12}$ and $\theta_{13}$, where
$c_{ij}\equiv\cos\theta_{ij}$ and $s_{ij}\equiv\sin\theta_{ij}$.\label{tab1}}
\begin{tabular}{|c||c|c|}\hline
$\phaa k\phaa $ &\phaa $q_{k1}\phaa $ & \phaa $q_{k2} \phaa $ \\ \hline
$1$ & $c_{12} c_{13}$ & $-s_{12}-ic_{12}s_{13}$ \\
$2$ & $s_{12} c_{13}$ & $c_{12}-is_{12}s_{13}$ \\
$3$ & $s_{13}$ & $ic_{13}$ \\
$4$ & $i$ & $0$ \\ \hline
\end{tabular}
\end{table}

\section{Theory: The Oblique Parameters $S$, $T$, and $U$ and Their Higher-Order Extensions $V$, $W$ and $X$}
The oblique parameters S, T, and U provide an indirect probe of physics beyond the SM for theories with $SU(2)\times U(1)$ gauge content.  They quantify deviations from the Standard Model in terms of radiative corrections to the gauge-boson two point functions, making use of the precise measurements available for parameters associated with W and Z boson resonances.  New physics contributions is encoded in $\delta \Pi_{ab}(q^2)$, where $a,b = \gamma, W^\pm, Z$ and
\beq \Pi_{ab}(q^2) = \Pi_{ab}^{SM}(q^2) +\delta\Pi_{ab}(q^2) .\eeq
The 2HDM is a good candidate for analysis in the oblique correction formalism because the theory has no new electroweak gauge boson content and its couplings to light fermions enter only at two loops and are thus suppressed relative to the gauge boson couplings. 

If the new physics enters at the TeV scale, the effect of the theory will be well-described by expansion to linear order in $q^2$, requiring only the three parameters (S, T, and U) originally defined by Peskin and Takeuchi \cite{peskin}:  
\beqa&&S=\left(\frac{4s_w^2c_w^2}{\alpha }\right)\left(\left[\frac{\PMzz-\Pozz}{\mz}\right]-\frac{\left(c_w^2-s_w^2\right)}{s_wc_w}\Popza -\Popaa\right),\label{s}\\
&&T=\left(\frac{1}{\alpha  }\right)\left[\frac{\Poww}{\mw}-\frac{\Pozz}{\mz}\right],\label{T}\\
&&U=\left(\frac{4s_w^2c_w^2}{\alpha }\right)\left(\left[\frac{\Pww-\Poww}{\mw}\right]-c_w^2 \left[\frac{\PMzz-\Pozz}{\mz}\right]\nonumber\right.\\
&&\biggl.\phantom{aaaa}-2c_w s_w\Popza -s_w^2\Popaa\biggr),\label{u}\eeqa
where $\Pi'(q^2) \equiv \frac{d}{dq^2}\Pi(q^2)$.  These expressions, reprinted here from Burgess et al. \cite{burgess93}, do not assume that the corrections vanish beyond linear order. If the vacuum polarization functions $\Pi_{ab}$ are expanded beyond linear order, six independent parameters are required to encode all the radiative corrections.  The three new parameters may be defined as follows~\cite{burgess93}:
\beqa 
&&V=\left(\frac{1}{\alpha  }\right)\left[\PMpzz-\frac{\PMzz-\Pozz}{\mz}\right],\label{v}\\
&&W=\left(\frac{1}{\alpha}\right)\left[\Ppww -\frac{\Pww-\Poww}{\mw}\right],\label{w}\\
&&X = \left(\frac{1}{\alpha}\right)\left(-s_w c_w\right)\left[\frac{\Pza-\Poza}{\mz}-\Popza\right].\label{x}\eeqa
These functions will be negligible if the new physics enters at a scale much larger than $m_Z$.  

\subsection{Calculating the Oblique Parameters in the 2HDM}\label{theoryobl}
Working on the assumption that the mass scale of the 2HDM is larger than $m_Z$ but not so high that V, W, and X are irrelevant, we present here the contributions to the extended oblique parameters in the basis-independent formalism. These results are based on the one-loop calculations in  refs.~\cite{diss} and \cite{haberoneil}, with the modification that contributions from the $A^\mu A_\mu \hp\hm$ and $A^\mu Z_\mu \hp\hm$ vertices have been added.  The one-loop diagrams constructed from these two vertices cancel in the oblique parameters, but they are included here to show explicitly that $\Poza$ and $\Poaa$ vanish as required~\cite{burgess93}. The required vacuum polarization functions are as follows:

\beqa&&\QMzz=\frac{\alpha }{4\pi  s_w{}^2c_w{}^2}\biggl(-\mz\left[q_{11}^2 B_0\left(q^2,\mz,m_1^2\right)+q_{21}^2B_0\left(q^2,\mz,m_2^2\right)+q_{31}^2B_0\left(q^2,\mz,m_3^2\right)\right]\nonumber\\
&&+q_{11}^2B_{22}\left(q^2,\mz,m_1^2\right)+q_{21}^2B_{22}\left(q^2,\mz,m_2^2\right)+q_{31}^2B_{22}\left(q^2,\mz,m_3^2\right)+q_{21}^2B_{22}\left(q^2,m_1^2,m_3^2\right)
\nonumber\\&&+q_{11}^2B_{22}\left(q^2,m_2^2,m_3^2\right)+q_{31}^2B_{22}\left(q^2,m_1^2,m_2^2\right)+c_{2w}{}^2B_{22}\left(q^2,\mcs,\mcs\right)-\frac{1}{2}[A_0\left(m_1^2\right)+A_0\left(m_2^2\right)
\nonumber\\&&\biggl.+A_0\left(m_3^2\right)+c_{2w}{}^2A_0\left(\mcs\right)]+\mz B_0\left(q^2,\mz,m_1^2\right)-B_{22}\left(q^2,\mz,m_1^2\right)\biggr),\eeqa

\beqa
&&\QMpzz=\frac{\alpha }{4\pi  s_w^2c_w^2}\biggl(-\mz\left[q_{11}^2 B_0'\left(q^2,\mz,m_1^2\right)+q_{21}^2B_0'\left(q^2,\mz,m_2^2\right)+q_{31}^2B_0'\left(q^2,\mz,m_3^2\right)\right]\biggr.\nonumber\\
&&+q_{11}^2B_{22}'\left(q^2,\mz,m_1^2\right)+q_{21}^2B_{22}'\left(q^2,\mz,m_2^2\right)+q_{31}^2B_{22}'\left(q^2,\mz,m_3^2\right)+q_{21}^2B_{22}'\left(q^2,m_1^2,m_3^2\right)\nonumber\\
&&+q_{11}^2B_{22}'\left(q^2,m_2^2,m_3^2\right)+q_{31}^2B_{22}'\left(q^2,m_1^2,m_2^2\right)+c_{2w}{}^2B_{22}'\left(q^2,\mcs,\mcs\right)+\mz B_0'\left(q^2,\mz,m_1^2\right)\nonumber\\
&&-B_{22}'\left(q^2,\mz,m_1^2\right)\biggr),\eeqa

\beqa && \Qpww=\frac{\alpha }{4\pi  s_w^2}\biggl(-\mw\left[q_{11}^2B_0'\left(q^2,\mw,m_1^2\right)+q_{21}^2B_0'\left(q^2,\mw,m_2^2\right)+q_{31}^2B_0'\left(q^2,\mw,m_3^2\right)\right]\biggr.\nonumber\\
&&+q_{11}^2B_{22}'\left(q^2,\mw,m_1^2\right)+q_{21}^2B_{22}'\left(q^2,\mw,m_2^2\right)+q_{31}^2B_{22}'\left(q^2,\mw,m_3^2\right)+q_{12}^2B_{22}'\left(q^2,\mcs,m_1^2\right)\nonumber\\
&&\biggl.+q_{22}^2B_{22}'\left(q^2,\mcs,m_2^2\right)+q_{32}^2B_{22}'\left(q^2,\mcs,m_3^2\right)+\mw B_0'\left(q^2,\mw,m_1^2\right)-B_{22}'\left(q^2,\mw,m_1^2\right)\biggr),\eeqa

\beqa && \Qww=\frac{\alpha }{4\pi  s_w^2}\biggl(-\mw\left[q_{11}^2B_0\left(q^2,\mw,m_1^2\right)+q_{21}^2B_0\left(q^2,\mw,m_2^2\right)+q_{31}^2B_0\left(q^2,\mw,m_3^2\right)\right]\biggr.\nonumber\\
&&+q_{11}^2B_{22}\left(q^2,\mw,m_1^2\right)+q_{21}^2B_{22}\left(q^2,\mw,m_2^2\right)+q_{31}^2B_{22}\left(q^2,\mw,m_3^2\right)+q_{12}^2B_{22}\left(q^2,\mcs,m_1^2\right)\nonumber\\
&&+q_{22}^2B_{22}\left(q^2,\mcs,m_2^2\right)+q_{32}^2B_{22}\left(q^2,\mcs,m_3^2\right)-\frac{1}{2}\left[A_0(m_1^2)+A_0(m_2^2)+A_0(m_3^2)+A_0(\mcs)\right]\nonumber\\
&&\biggl.+\mw B_0\left(q^2,\mw,m_1^2\right)-B_{22}\left(q^2,\mw,m_1^2\right)\biggr),\eeqa

\beq \Qaa=\frac{\alpha }{\pi  }\left[B_{22}\left(q^2,\mcs,\mcs\right)-\half A_0\left(\mcs\right)\right],\eeq

\beq \Qpaa=\frac{\alpha }{\pi}\left[B_{22}'\left(q^2,\mcs,\mcs\right)\right],\eeq

\beq \Qza=\frac{\alpha }{2\pi  s_w}\frac{c_{2w}}{c_w}\left[B_{22}\left(q^2,\mcs,\mcs\right)-\half A_0\left(\mcs\right)\right].\eeq

The following functions vanish, as required by the Ward identities, and thus may be removed from \eq{x} without loss of generality:

\beqa && \Poza=\frac{\alpha }{2\pi  s_w}\frac{c_{2w}}{c_w}\left(B_{22}\left[0,\mcs,\mcs\right]\right)-\half A_0\left(\mcs\right)=0,\nonumber\\
&&\Poaa=\frac{\alpha }{\pi}\left[B_{22}\left(0,\mcs,\mcs\right)\right]-\half A_0\left(\mcs\right)=0.\eeqa

%%%%%%%%%%%%%%% NUMERICAL RESULTS
\section{What the 2HDM Predicts for the Oblique Parameters}

\subsection{Comparing with Experimental Limits}\label{sec:hist}
In this section we compare the theoretical results for the oblique parameters with those of experiment.  The experimental limits for S, T, and U are given in \Ref{erlerPDG}: 

\beqa S &=&.014\pm .10,\\
T &=& .03\pm .11,\\
U &=& .06\pm .10.\label{pdglim}
\eeqa

The limits for V, W, and X are known for a reference mass of 100 GeV \Ref{kr97}:
\beqa W &=& .11\pm 4.7,\\
V &=& .30\pm .38,\\
X &=& .38 \pm .59.\label{wvxlim}\eeqa

To compare the experimental limits with the values predicted by the 2HDM, we randomly sampled the parameter space of the scalar sector, choosing arbitrary values of\footnote{We  use $\zthf$ as a parameter rather than $\zf$ since $\zf$ is not by itself constrained by unitarity.} $\zone$, $\zth$, $\zthf$, $\zfiver$, $\zsixr$, $\zfii$ and $\zsixi$, within the unitarity limits calculated in \Ref{haberoneil}.  Each set of $Z_i$ parameters [$\zone$, etc.] determines the scalar masses and the values of the $q_{k\ell}$ functions, producing unique values of S, T, U, V, W, and X via the expressions in Section~\ref{theoryobl}.  Since $Y_2$ is not constrained by unitarity, $Y_2 = m_W^2$ was used for all points.  Although there is not a definite lower bound from experiment for $m_1$ as there is for the Standard Model Higgs mass, our code discards points that generate  scalar masses less than $m_Z$.\footnote{Because the 2HDM produces very few points at low mass scales, discarding these points has an insignificant effect on the results.  This subject will be discussed in more detail in Section~\ref{constraint}.} This cut-off was chosen so that our code would produce sensible results regardless of whether the oblique parameters are defined in the linear/quadratic approximation or defined to all orders in the $q^2$ expansion.

\begin{figure}[h!]
\centering
\begin{tabular}{cc}
\includegraphics[width=.4\linewidth]{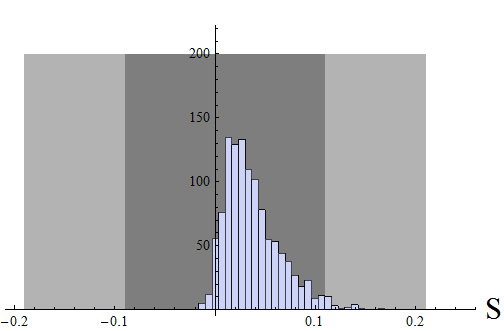}
&\hspace{4mm}
\includegraphics[width=.4\linewidth]{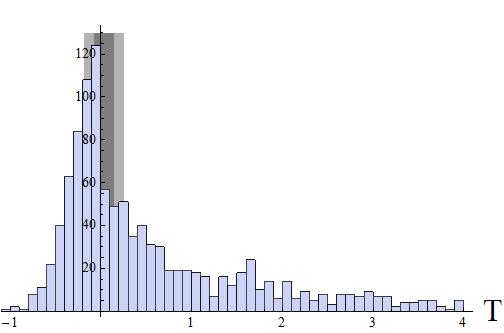}
\\
\includegraphics[width=.4\linewidth]{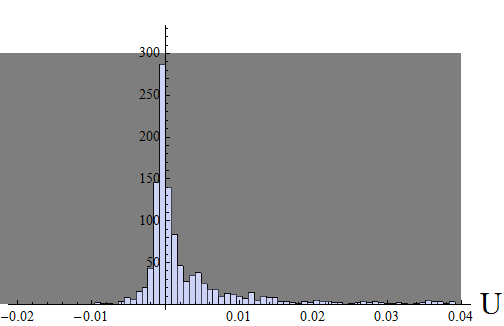}
&\hspace{4mm}
\includegraphics[width=.4\linewidth]{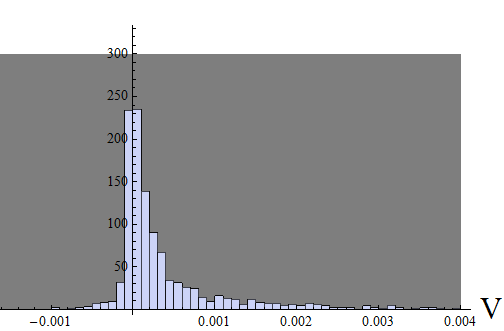}
\\
\includegraphics[width=.4\linewidth]{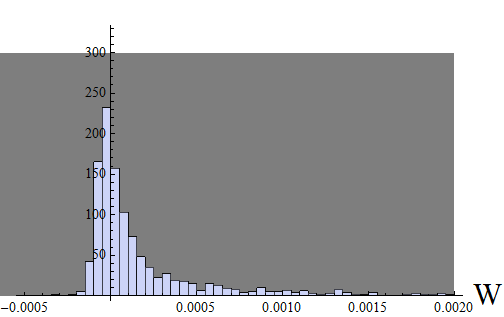}
&\hspace{4mm}
\includegraphics[width=.4\linewidth]{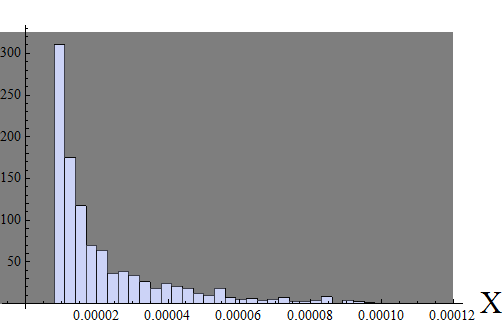}
\end{tabular}\label{oblhist}\caption{Histograms showing the distribution of S, T, U, V, W and X values generated by a random sampling of 2HDM parameter space.  The experimental bounds are shown as light gray shading (2$\sigma$ limits) and dark gray shading (1$\sigma$ limits).}
\end{figure} 

The values of the oblique parameters generated in this random sample of 2HDM parameter space are shown in histogram form in Fig.~\ref{oblhist}.  With the exception of the T parameter, they fall well within the experimental bounds.  Because the 2HDM generically produces values of T that are unrealistically large in magnitude, much of the parameter space will be eliminated on experimental grounds.  This result will be explored further in subsequent sections. On the other hand, the distributions for S, U, V, W, and X fall well within the experimental limits, and thus are not useful in either constraining the parameter space or ruling out the 2HDM.  Furthermore, these values all fall within roughly .1 of zero, implying that the 2HDM may be indistinguishable from the SM on the basis of these parameters.

From the histograms, it is evident that the 2HDM predicts positive values for S and X.  Since a random scan of parameter space tends to produce masses well above $m_Z$, the values for V, W, and X tend to be at least an order of magnitude smaller than for S, T, and U (more than one order of magnitude in the cases of W and X).  This suggests that the 2HDM corrections are well described by S, T, and U; one does not gain much by going beyond the linear expansion approximation.  These results are not specific to the particular value of $Y_2$ chosen; the allowed ranges of the $Z_i$ parameters (determined by unitarity) have more effect than the specific choice of $Y_2$ on the oblique parameters as long as $Y_2$ is less than or on the order of the square of the electroweak scale.   Thus, choosing a smaller value for $Y_2$ does not produce larger ranges of the oblique parameters.  

\begin{figure}[h!]
\includegraphics[width=.4\linewidth]{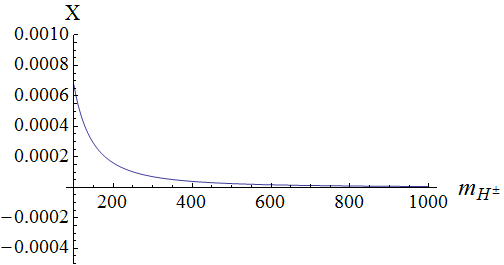}
\caption{To first order, the parameter X is proportional to $\mz/\mc^2$.}
\label{xm}
\end{figure}

The histogram for the X parameter has a different form from the others and merits some comment.  In the 2HDM, X is a function of $\mc$ only, as illustrated in Fig.~\ref{xm}.  To linear order in $\mz/\mc^2$ it is equal to

\beq X = \frac{\ctw^2}{4 \pi}\cdot\frac{1}{60}\cdot\frac{\mz}{\mc^2} + O(\frac{m_Z^4}{\mc^4}).\label{xanalytic} \eeq

Since X is proportional to $1/\mc^2$, its lower bound depends on the maximum value of $\mc$, which is controlled by $\zth$ and $Y_2$.  This is reflected in the histogram; the location of the sharp cut-off derives from the constraints from unitarity on the parameter $\zth$; if the unitarity bounds are relaxed, or if a higher value of $Y_2$ is chosen, the upper limit on $\mc$ increases, shifting the histogram data to the left.

Other than X, the parameters fall into roughly bell-shaped distributions around their most probable values, exhibiting longer tails on the positive side than on the negative side.  The statistics for these parameters are summarized in Table~\ref{stat}.

\begin{table}[h!]
\centering
\begin{tabular}{|c|c|c|c|}\hline
 &Central Value& Range & Standard Deviation\\\hline
S&0.030&(-0.016, 0.19)&0.029\\
T&0.19&(-1.2, 8.5)&1.6\\
U&0.00015&(-0.010, 0.060)&0.0088\\
V&0.00011&(-0.0010, 0.014)&0.0013\\
W&0.000033&(-0.00033, 0.0086)&0.00080\\
X&0.000016&($8.3\cdot10^{-6}$, 0.00073)&0.000066\\\hline
\end{tabular}\caption{Predictions of the CP-violating 2HDM: Statistics for 1000 randomly produced points, with $Y_2 = \mw$. }\label{stat}
\end{table} 

\begin{figure}[h!]
\centering
\begin{tabular}{cc}
\includegraphics[width=.4\linewidth]{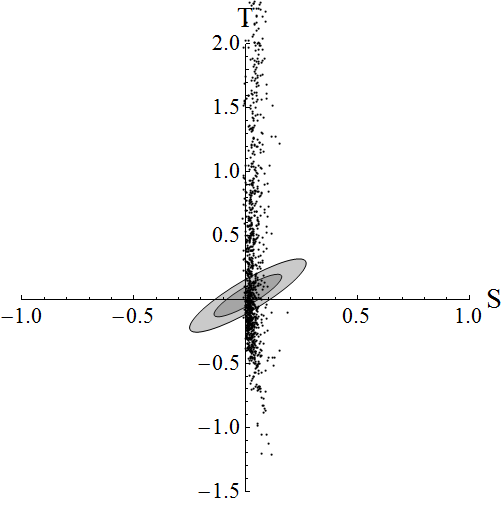}
&\hspace{4mm}
\includegraphics[width=.4\linewidth]{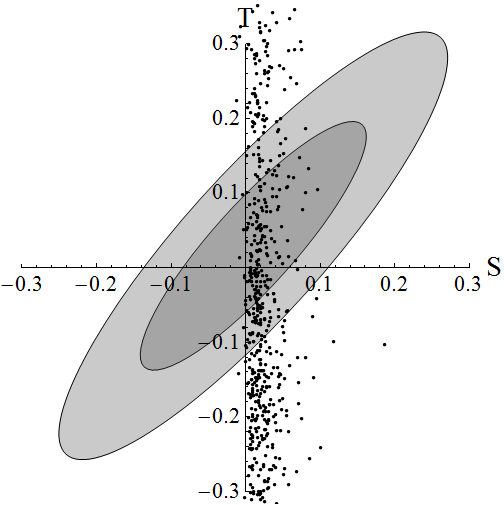}
\end{tabular}\caption{Scatterplots of randomly generated points superposed on experimental limits in S-T space.  The light gray shading shows the $1~\sigma$ bounds and the dark gray shading shows the $2~\sigma$ bounds. The second figure zooms in on the allowed region.}\label{ell}
\end{figure}

\begin{figure}[h!]
\centering
\begin{tabular}{cc}
\includegraphics[width=.4\linewidth]{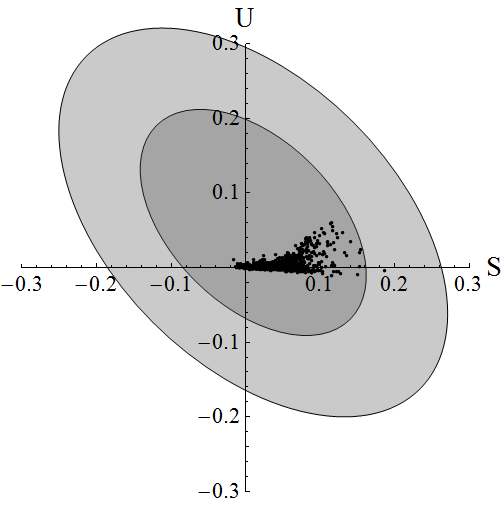}
&\hspace{4mm}\includegraphics[width=.4\linewidth]{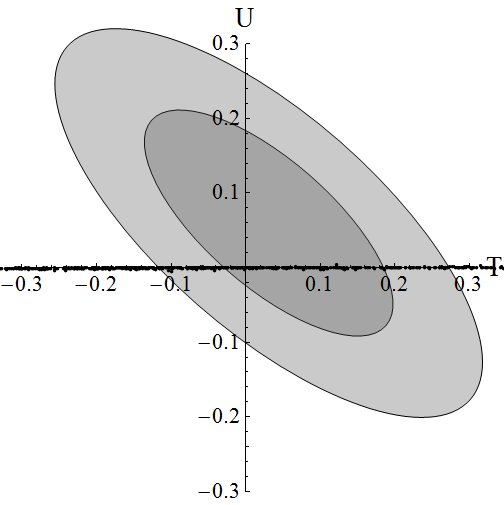}
\end{tabular}\caption{Scatterplots of randomly generated points superposed on experimental limits in S-U space [left] and  T-U space [right].  The light gray shading shows the $1~\sigma$ bounds and the dark gray shading shows the $2~\sigma$ bounds.}\label{ell2}
\end{figure}

Because the experimental limits for each oblique parameter are not independent from one other, it is conventional to display results using plots of allowed ellipses in S-T space (or other 2D parameter spaces).  The OPUCEM library~\cite{opucem} can be used for this task.  We adapted the OPUCEM algorithms for use in Mathematica to produce ellipses superposed on our scatterplots of data.\footnote{Our Mathematica program for generating the elliptical contours can be downloaded from the following site: http://people.bridgewater.edu/\~{}doneil/}    The result for the S-T ellipse is shown in Fig.~\ref{ell}; ellipses in S-U and T-U space are shown in Fig.~\ref{ell2}.  These plots show that the points generated by the 2HDM are consistent with the experimental limits (with the exception of T, as noted previously).

\begin{figure}[h!]
\centering
\begin{tabular}{cc}
\includegraphics[width=.4\linewidth]{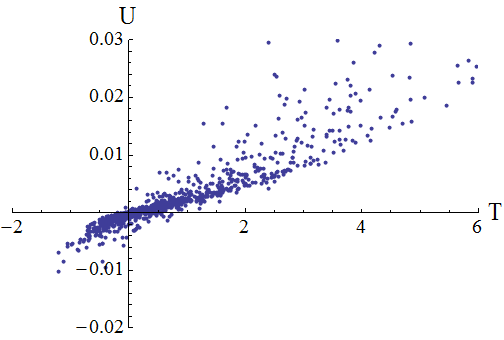}
&\hspace{4mm}
\includegraphics[width=.4\linewidth]{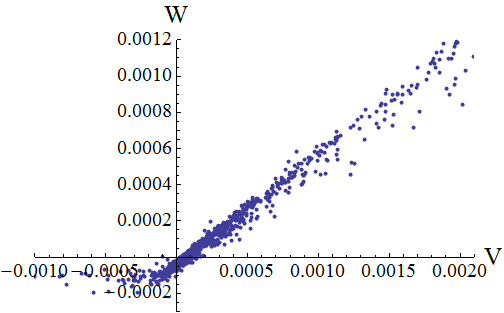}
\end{tabular}\caption{Correlations appear in two sets of the oblique parameter values, T-U and V-W.  The rest of the oblique parameters (not shown) do not appear to exhibit strong correlations with each other.}\label{corr}
\end{figure}

The points with larger values of U that are evident in the S-U ellipse are not evident in the T-U ellipse because there is a strong correlation in the 2HDM between U and T.  Values of U that are larger than .01 correspond to T $>1$, and thus do not appear in Fig.~\ref{ell2}.  Incidentally, there is another strong correlation that crops up in the 2HDM: a near-linear relationship between V and W in the first quadrant.  These correlations are illustrated in Fig.~\ref{corr}.

These two relationships probably have a similar origin.  For example, examining the definitions in \eqs{v}{w}, one notes that V (which is based on ZZ diagrams) is defined analogously to W (based on WW diagrams), so the fact that the vacuum polarization functions $\Pi_{WW}(q^2)$ and $\Pi_{ZZ}(q^2)$ have many similar terms leads V and W to have similar forms. To a lesser extent, the same phenomenon occurs with T and U.  As a result of these correlations, the actual constraints on the 2HDM are more strict than those implied by the individual distributions illustrated in Figs.~\ref{oblhist}.

%%%%%%%%%% KUNDU AND ROY ANALYSIS

\section{Other Definitions of the Oblique Parameters and the Effect of the $q^2$-Expansion Approximation}
Two distinct definitions of the oblique parameters appear in the literature; those of Peskin and Takeuchi~\cite{peskin}, and the similar but physicially inequivalent expressions of Marciano and Rosner~\cite{mr}.  For the purpose of numerically analyzing the oblique parameters in the 2HDM, we have used in this paper the definitions of S, T, and U of Burgess et al.~\cite{burgess93}, which are generalizations of the MR definitions.  They do not assume that the vacuum polarization functions $\Pi(q^2)$ are well-described by an expansion linear in $q^2$.   Kundu and Roy~\cite{kr97} advocate alternative expressions for S and U that also do not rely on the linear expansion approximation:

\beqa&&S=\left(\frac{4s_w^2c_w^2}{\alpha }\right)\left(\left[\frac{\PMzz-\Pozz}{\mz}\right]-\frac{\left(c_w^2-s_w^2\right)}{s_wc_w}\left[\frac{\Pza}{\mz}\right] -\left[\frac{\Paa}{\mz}\right]\right),\label{sroy}\\
&&U=\left(\frac{4s_w^2c_w^2}{\alpha }\right)\left(\left[\frac{\Pww-\Poww}{\mw}\right]-c_w^2 \left[\frac{\PMzz-\Pozz}{\mz}\right]\nonumber\right.\\
&&\biggl.\phantom{aaaa}-2c_w s_w\left[\frac{\Pza}{\mz}\right] -s_w^2\left[\frac{\Paa}{\mz}\right]\biggr),\label{uroy}\eeqa

\begin{figure}[h!]
\centering
\begin{tabular}{cc}
\includegraphics[width=.4\linewidth]{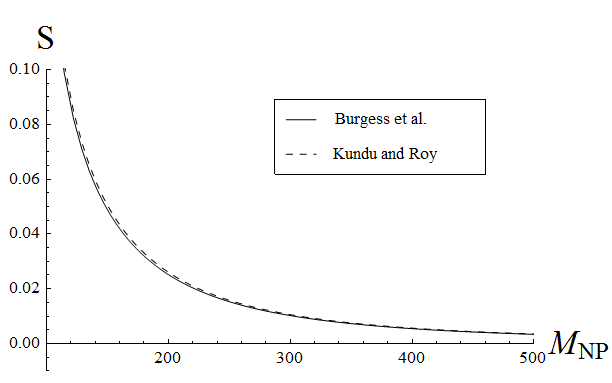}
&\hspace{4mm}
\includegraphics[width=.4\linewidth]{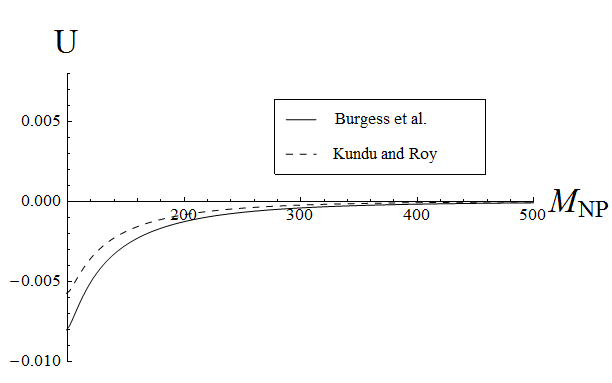}
\end{tabular}\caption{S and U, defined by Burgess et al. (solid line) and  Kundu and Roy (dashed line).  The two should be physically equivalent in the limit $q^2 \ll M_{NP}^2$; hence the convergence as $M_{NP}$ increases.}\label{linfig}
\end{figure}

Since these expressions are based on the PT definitions, they have a slightly different physical interpretation than those of \eqs{s}{u}.  To linear order in $q^2$, there is no difference between the two, but beyond linear order they differ by a factor on the order of the X parameter~\cite{kr97}.  To assure the reader that no significant errors are introduced by applying the PDG experimental limits (based on the PT definitions) to theoretical computations in the Burgess scheme, comparisons of the two sets of oblique parameters are shown in Fig.~\ref{linfig}. To examine the behavior of the functions S and U as the mass scale of the new physics (NP) increases, we define
\beq M_{NP}\equiv \quarter \left(m_1 + m_2 + m_3+\mc\right),\label{massav}\eeq
and choose
\beqa 
&\zone = .5~( \zth) + .5,~~~~Y_2 = \mw, ~~~~\zf = 0,\nonumber\\
& \zfiver = \zsixr= -.1, ~~~~\zfii =  \zsixi = .1.\label{increasemass}\eeqa
With these values for the scalar couplings, dialing up $\zth$ simultaneously increases the neutral masses $m_i$ and the charged Higgs mass $\mc$.  Thus, one can analyze the dependence of the oblique parameters on the mass scale $M_{NP}$ by varying $\zth$.  The results in Fig.~\ref{linfig} show that the two definitions converge in the limit $q^2\ll M_{NP}^2$, as they should.  Even at lower mass scales the difference is insignificant for S.  There is noticeable difference at low $M_{NP}$ for U, but the difference is dwarfed by the experimental error in U [see \eq{pdglim}], so this should not undermine confidence in our numerical results.

Meanwhile, there are also different ways of calculating W and V.  The definitions advocated by Kundu and Roy in \Ref{kr97} are mathematically equivalent to those introduced by Burgess et al. [see \eqs{v}{w}], so the experimental limits in \eq{wvxlim} apply equally well to either.  For completeness, one can also check that these expressions for W and V converge at high mass scales to those defined in the quadratic approximation:
\beq \Pi(q^2) = \Pi(0) + q^2 \Pi'(0) + \half q^4 \Pi''(0).\label{quad}\eeq
In this approximation, V and W become:
\beqa 
&&V=\left(\frac{1}{\alpha  }\right)\left[\frac{\PMzz-\Pozz}{\mz}-\Popzz\right],\\
&&W=\left(\frac{1}{\alpha}\right)\left[\frac{\Pww-\Poww}{\mw} -\Popww\right].\eeqa
In reproducing these equations from \Ref{kr97} we have converted them to our notation, in which $\Pi'(q^2) \equiv \frac{d}{dq^2}\Pi(q^2)$.  Figure~\ref{quad} show that V and W in this quadratic approximation converge to the non-approximate forms as the mass scale of the new physics increases, as expected.

\begin{figure}[h!]
\centering
\begin{tabular}{cc}
\includegraphics[width=.4\linewidth]{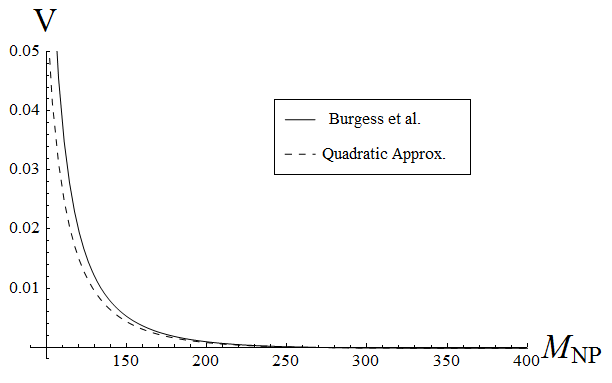}
&\hspace{4mm}
\includegraphics[width=.4\linewidth]{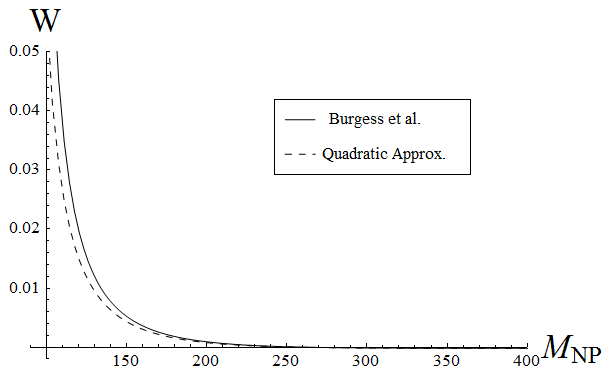}
\end{tabular}\caption{Comparison of V and W in the formulation by Burgess et al. and in the quadratic approximation.}\label{quad}
\end{figure} 

%%%%%%%%%%%%%%%%%%%%%%%%%%%%%%%%%%%%%%%%%%%%%%%%%
\section{Constraining the 2HDM with the Oblique Parameters and Unitarity Bounds}\label{constraint}
\subsection{Constraints on the Mass of the Charged Higgs Boson}
\begin{figure}[h!]
\centering

\includegraphics[width=.4\linewidth]{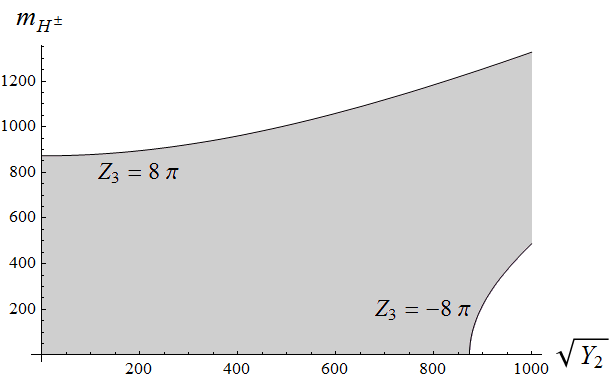}

\caption{Range of $\mc$ values allowed by the unitarity limits on $Z_3$, as a function of $\sqrt{Y_2}$ (in GeV).  For values of $\sqrt{Y_2}$ around the electroweak scale, the upper limit on $\mc$ is about $900$ GeV.}\label{y2m}
\end{figure}

In this section we discuss what the oblique parameters indicate about the charged Higgs mass $\mc$.  It follows from \eq{mch} that the unitarity bound $|\zth|<8 \pi$ affects $\mc$.  In Fig.~\ref{y2m}, the allowed region for $\mc$ is shown.  Without knowing $Y_2$, one cannot put a definite upper bound on $\mc$.  However, by using the experimental limits on the oblique parameters, a more precise bound may be obtained.

The most promising of the oblique parameters in this regard are S and T.  For the purposes of this analysis, we will use the PDG values of S and T with U fixed to be zero, since most values of U were found to be within .01 of zero in section~\ref{sec:hist}:
\beqa S &=& 0.03 \pm 0.09,\nonumber\\
T &=& 0.07 \pm 0.08.\label{fixedST}\eeqa

\begin{figure}[h!]
\centering
\begin{tabular}{cc}
\includegraphics[width=.4\linewidth]{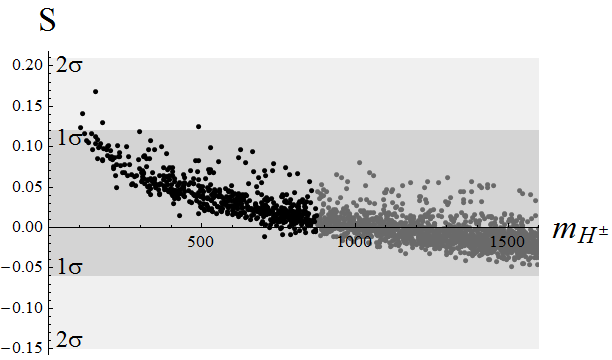}
&\hspace{4mm}
\includegraphics[width=.4\linewidth]{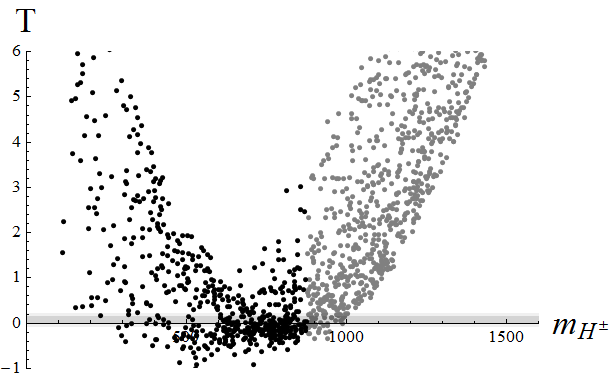}
\end{tabular}\caption{Scatterplots of randomly generated S and T values, with $Y_2 = \mw$. The black points are allowed; the gray points violate the unitarity limits on $\zth$.  The experimental values are shown as a dashed line; the gray shading shows the 1$\sigma$ and 2$\sigma$ error bars.}\label{unit}
\end{figure}

The behavior of S and T as a function of $\mc$ is shown in Fig.~\ref{unit}.  The points in the scatterplots were generated as in section~\ref{sec:hist}, with the added inclusion of extra points (shown in gray) that appear when the unitarity constraint on $\zth$ is relaxed by a factor 4 (ie, $|\zth|<32 \pi$ instead of $|\zth|<8 \pi$). The plots show that low values of S are associated with higher values of $\mc$.  The distribution of S values predicted by the 2HDM (according to randomly selecting the scalar parameters within the unitarity limits) has a peak around S = .02 (a value that agrees well with experiment), which correlates to a relatively high value of charged Higgs mass ($\mc > 600~$GeV).  The results for T also appear to disfavor low values of $\mc$; although points producing $\mc \lsim 250~$GeV are easily generated, they tend to produce unrealistically high values of T.  It appears, then, that the 2HDM predicts
\beq \mc \gsim 250~ \rm{ GeV},\label{mcpred}\eeq
with higher values favored.

\begin{figure}[h!]
\centering
\begin{tabular}{cc}
\includegraphics[width=.4\linewidth]{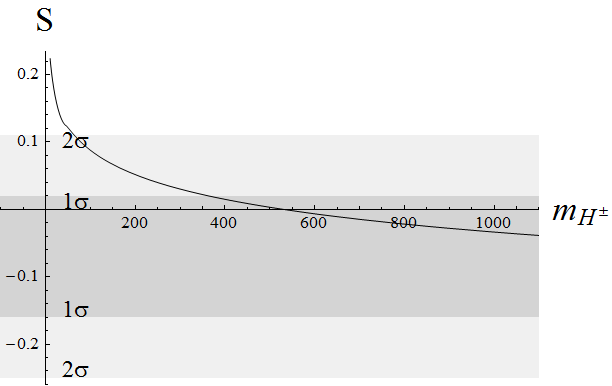}
&\hspace{4mm}\includegraphics[width=.4\linewidth]{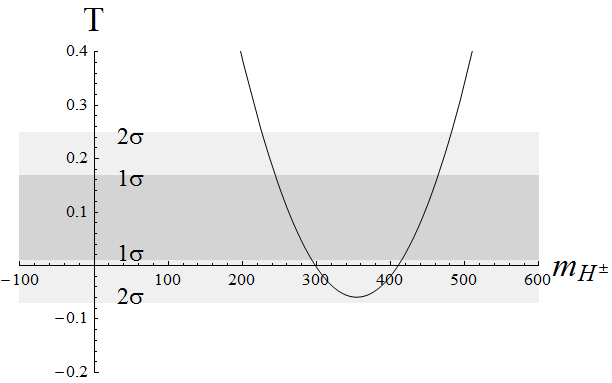}
\label{onepoint}
\end{tabular}
\caption{The behavior of S and T at the point given by \eq{specialhigh}, where $m_1 = 321~ \rm{GeV}.$ With this value as a reference mass, the experimental limits shift slightly [see \eq{higher}]. }
\end{figure}

One should be careful in interpreting the results in Fig.~\ref{unit}.  For one thing, the experimental limits in \eq{fixedST} assume a Standard Model Higgs mass of $m_\phi =  117$ GeV.  In our analysis, we have been fixing $m_\phi = m_1$.  Since $m_1$ does not equal 117 GeV in general in this analysis, the experimental bounds will not be completely accurate for all points.  To determine whether this is significant,  let us examine the S and T scatterplots in more detail.  Essentially what appear in the scatterplots are sets of possible curves, which exceed the experimental bounds at different points. One such curve is shown in Fig.~\ref{onepoint}, generated by choosing
\beqa 
&\zone = 3.5,~~~~Y_2 = \mw, ~~~~\zthf = 4,\nonumber\\
&  \zfii = \zfiver = \zsixr= \zsixi = .5,\label{specialhigh}\eeqa
with varying values of $\zth$ to produce a function of $\mc$.

The neutral Higgs boson masses for this choice of parameters are
\beq m_1 = 321~ \rm{GeV},~~~~ m_2 = 363 ~\rm{GeV}, ~~~~m_3 = 483~\rm{GeV}.\eeq
For $m_\phi =  300$ GeV, the PDG gives different experimental limits:
\beqa S &=& -0.07 \pm 0.09,\nonumber\\
T &=& 0.09 \pm 0.08\label{higher},\eeqa
so the limits on the plots have been adjusted accordingly [see Fig.~\ref{onepoint}]. Thus, it does not appear that shifting the experimental limits to correspond to a higher SM reference mass appreciably affects the conclusion of \eq{mcpred}.

However, one more caveat is required, which we will illustrate as follows:  Consider the following parameter values,  \beqa 
&\zone = 3.5,~~~~Y_2 = \mw, ~~~~\zthf = 1,\nonumber\\
&  \zfii = \zfiver = \zsixr= \zsixi = .5,\label{specialpt},\eeqa
which generate the following mass values:
\beq m_1 = 117~ \rm{GeV},~~~~ m_2 = 221 ~\rm{GeV}, ~~~~m_3 = 472~\rm{GeV}.\eeq
\begin{figure}[h!]
\centering
\begin{tabular}{cc}
\includegraphics[width=.4\linewidth]{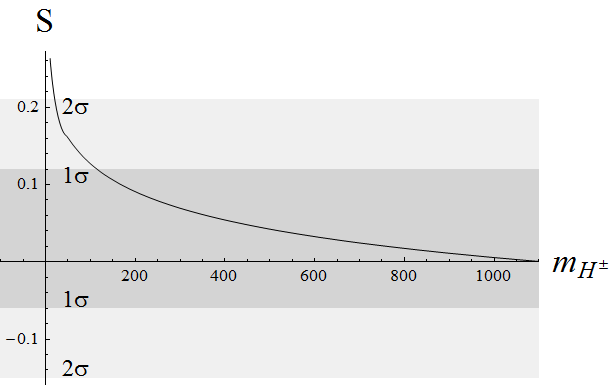}
&\hspace{4mm}
\includegraphics[width=.4\linewidth]{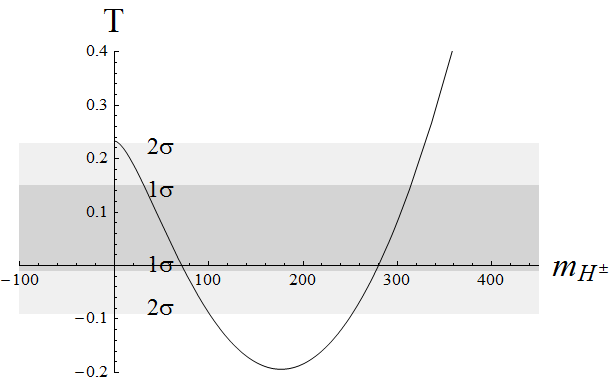}
\end{tabular}
\caption{The behavior of S and T at the point given by \eq{specialpt}, where $m_1 = 117~ \rm{GeV}.$ }\label{specialgr}
\end{figure}

\begin{figure}[h!]
\includegraphics[width=.8\linewidth]{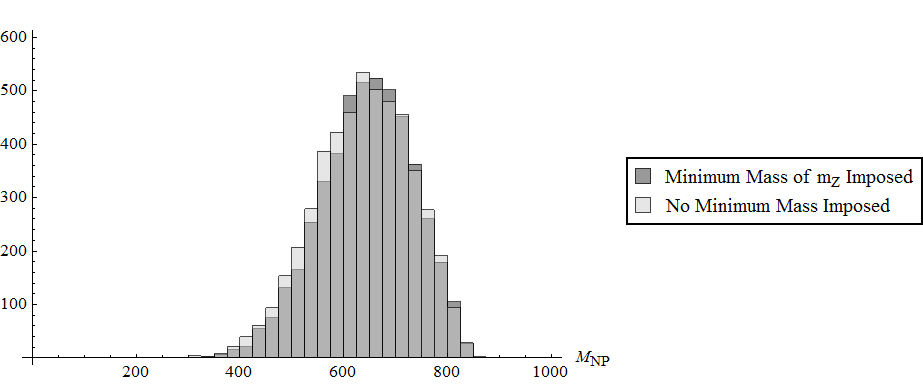}
\caption{Mass scales that result from a random sampling of 2HDM parameter space at $Y_2 = m_W^2$.  The mass scale $M_{NP}$ is the average of the three neutral masses and $\mc$ [see \eq{massav}].  The cut-off at the high end is due to imposing unitarity limits.  The strong overlap between the distributions shows that the lower limit is \emph{not} due to imposing $m_i \geq m_Z$ (as we have done throughout this analysis).}
\label{mass}
\end{figure}

The behavior of S and T at this point in parameter space is shown in Fig.~\ref{specialgr}. Note that in this figure the T curve has shifted to the left, into a region not represented on the scatterplots of Fig.~\ref{unit}.  In other words, the lower bound in \eq{mcpred} may be too restrictive, since it represents mostly higher sets of neutral Higgs masses.  Thus, while one can conclude that the 2HDM \textit{generically} predicts values of $\mc$ that are above 250 GeV, additional information about the theory may render the constraints from Fig.~\ref{unit} inapplicable.  (An example would be if the lightest Higgs mass was detected below 150 GeV.)  It should be noted, however, that the point defined in \eq{specialpt} is particularly nongeneric; it was tuned to give an especially low $m_1$ value. In general, the 2HDM predicts mass scales that are fairly high (400 - 800 GeV), as is evident from the distributions shown in Fig~\ref{mass}. To demonstrate that this favoring of high scales is not an artifact of our choice to discard points with $m_1 < m_Z$, the actual distribution produced by our program is compared in the figure to one with all real-valued masses allowed.  The figure shows that imposing this minimum mass condition introduces only a very slight biasing in the high-mass direction.

A final comment on the charged Higgs mass:  If the parameter X was known more precisely from experiment, it would be possible to determine $\mc$ from X alone.  X, as can be seen from \eq{xanalytic}, is an analytic function of $\mc$ and does not depend on any other scalar masses or couplings.  Unfortunately, the experimental bounds are 2 orders of magnitude too large to be useful in determining $\mc$ (see  Fig. \ref{xm}).

\subsection{Constraints on the Splitting between the Charged and Neutral Scalar Masses}
The difference between the charged Higgs mass ($\mc$) and the heaviest neutral mass ($m_3$) in the 2HDM is strongly constrained by the experimental bounds on the T parameter.  In Fig.~\ref{specialhigh}, the behavior of T at the point defined in \eq{specialhigh} is shown.  The experimental bounds on T constrain $m_3$ to be higher than $\mc$.

\begin{figure}[h!]
\centering
\includegraphics[width=.4\linewidth]{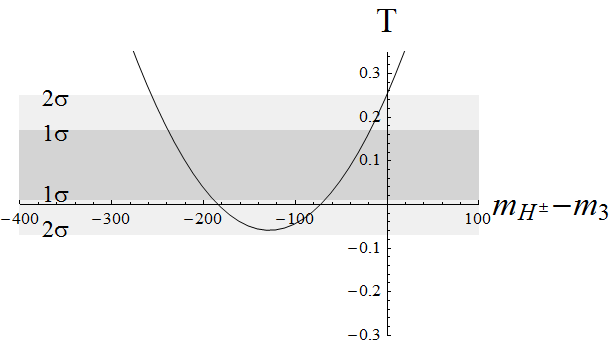}
\caption{The T parameter at the point given by at the point given by \eq{specialhigh}, with $m_1 = 321$ GeV. The gray shading shows the 1 $\sigma$ and 2 $\sigma$ error bars.  The experimental limits are shifted to match a 300 GeV reference mass, as in \eq{higher}.}\label{splspechigh}
\end{figure}
\begin{figure}[h!]
\centering
\begin{tabular}{cc}
\includegraphics[width=.4\linewidth]{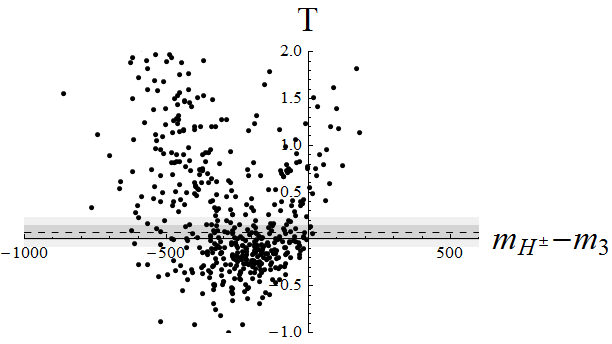}
&\hspace{4mm}
\includegraphics[width=.4\linewidth]{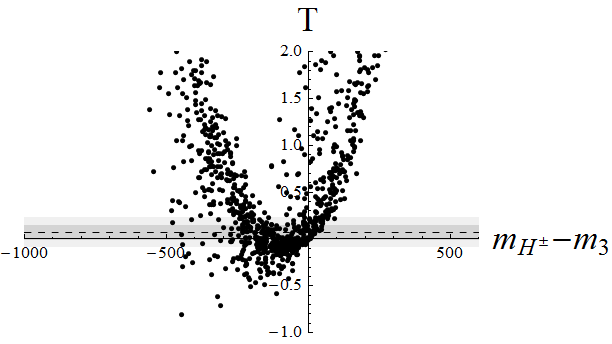}
\end{tabular}\caption{Scatterplots of randomly generated T values, for $Y_2 = (m_W)^2$ [left] and $Y_2 = (1 \rm{ TeV})^2$ [right]. The gray shading shows the 1 $\sigma$ and 2 $\sigma$ error bars.}\label{spl}
\end{figure}

 
In fact, this behavior is fairly generic--one can see from the scatterplots in Fig. \ref{spl} that for $ Y_2 = m_W^2$, the 2HDM seems to strongly favor $ \mc < m_3$.  In particular, the only points within the $2\sigma$ bounds on T are those with $\mc - m_3$ between $-600$  and $+50$ GeV.  

The role of T in constraining the splitting becomes especially important for higher values of $Y_2$.  For $Y_2 = (1\rm{ TeV})^2$, there are plenty of points for which $\mc > m_3$, but only those for which $\mc$ is less than $100$ GeV greater than $m_3$ survive the experimental constraints on T (see righthand image in Fig~\ref{spl}).  Thus, allowing for a wide variation in the possible values of $Y_2$, one can conclude from this analysis that
\beq
 -600  \rm{~GeV}<  \mc - m_3 < 100 \rm{~GeV}.\label{splitconstrain}\eeq

\begin{figure}[h!]
\centering
\includegraphics[width=.4\linewidth]{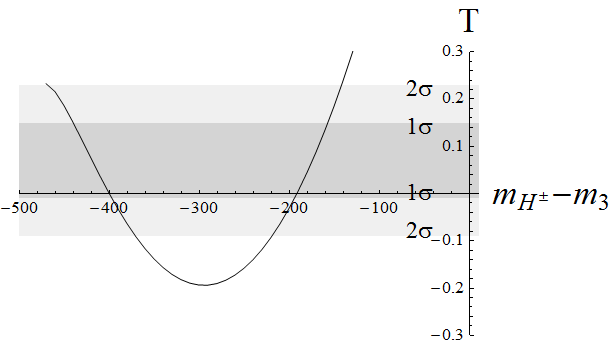}
\caption{The behavior of T at the point given by \eq{specialpt}, with $m_1 = 117$ GeV. }\label{splspec}
\end{figure}

Again, one should check that this conclusion remains valid for a light value of $m_1$.  Returning to the special point defined in \eq{specialpt}, with $m_1 = 117$ GeV, one sees from Fig.~\ref{splspec} that the curve has shifted entirely into the $m_3 > \mc$ region.
This is consistent with the result in \eq{splitconstrain}.

\section{The CPC model vs. the CPV model: Do they make different predictions for the oblique parameters?}
\begin{figure}[h!]
\centering
\begin{tabular}{cc}
\includegraphics[width=.45\linewidth]{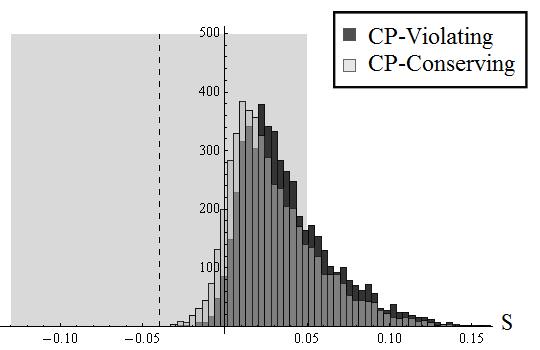}
&\hspace{4mm}
\includegraphics[width=.45\linewidth]{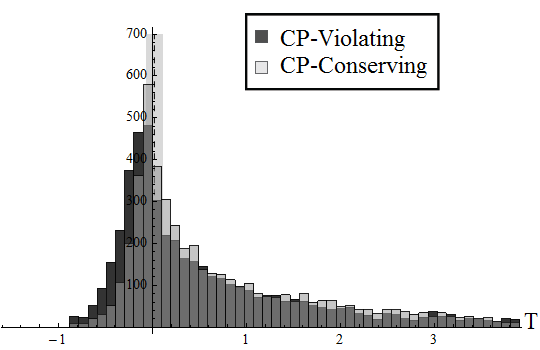}\end{tabular}
\caption{Taking the CP-conserving limit shifts the distribution of S values slightly in the negative direction [left] and shifts the distribution of T values slightly in the positive direction [right].  The experimental value of S and T are shown as dashed lines; the light gray shading shows the 1 $\sigma$ error bars.}
\label{fig:cp}

\end{figure}

The preceding analysis was done in the CP-Violating 2HDM.  To see if the results would change if the scalar sector of the model conserved CP, the oblique parameters were calculated in the CP-conserving limit, taking $\zfii = \zsixi = 0$.  This produced small shifts in the distribution of the oblique parameters; two examples of which are shown in Fig.~\ref{fig:cp}.  
The differences between the CP-violating and CP-conserving results were in all cases much smaller than the width of the distributions, confirming that the numerical findings in this paper apply to the CP-conserving 2HDM as well.  Unfortunately, this result precludes the possibility of predicting from experimental values of the oblique parameters whether the 2HDM would have CP-violation in the scalar sector.

\section{Conclusion}
This numerical analysis of the CP-violating 2HDM shows that the model generically predicts large scalar masses, with average masses in the 400-600 GeV range.  It is unsurprising, then, that a random sampling of the parameter space produces values of the extended oblique parameters W, V, and X that are at least an order of magnitude smaller than those of the traditional oblique parameters S, T, and U, since the extended oblique parameters are expected to be small except where the mass scale of the new physics is very close to the Z boson mass.  Larger values of W, V, and X are consistent with the 2HDM, but are not characteristic of the model.  We also find that the 2HDM produces values of U within .01 of zero, suggesting that the behavior of S and T is best compared to the experimental values where U has been fixed to be zero.  
The $T$ parameter, for which the 2HDM produces a wide range of values, puts constraints on the possible values of the charged Higgs mass and even more so on the splitting between the heaviest neutral scalar mass and the charged Higgs mass.  The 2HDM favors large $\mc$ values, of at least 250 GeV, and even larger values of $m_3$, the heaviest neutral Higgs boson.  In particular, we find $-600  \rm{~GeV}<  \mc - m_3 < 100 $ GeV.  This suggests that if a light scalar particle (say $\sim 120$ GeV) is discovered at the LHC, the possibility remains of heavier scalars being found closer to the TeV scale, with the charged Higgs particle expected to be lower in mass than the heaviest neutral state.  

These numerical results are not significantly affected by altering the definitions used for the oblique parameters, nor by going to the CP-conserving limit.  
\section{Acknowledgements}
The authors would like to thank Howard Haber for his generous help with this project.  Our work has also benefited from information provided by Michael Dine and Probir Roy.

This research was supported in part by the NSF-REU grant DMR-0649112.
%%%%%%%%%%%%%%%%%%%%%%%%%% BIBLIOGRAPHY   %%%%%%%%%%%%%%%%%%%%%%%%%%

\end{document}